# Direction-Reversing Quasi-Random Rumor Spreading with Restarts

Carola Winzen[*]

Max-Planck-Institut für Informatik, 66123 Saarbrücken, Germany

## Abstract

In a recent work, Doerr and Fouz [*Asymptotically Optimal Randomized Rumor Spreading*, in ArXiv] present a new quasi-random PUSH algorithm for the rumor spreading problem (also known as gossip spreading or message propagation problem). Their *hybrid protocol* outperforms all known PUSH protocols.

In this work, we add to the hybrid protocol a direction-reversing element. We show that this *direction-reversing quasi-random rumor spreading protocol with random restarts* yields a constant factor improvement over the hybrid model, if we allow the same dose of randomness.

Put differently, our protocol achieves the same broadcasting time as the hybrid model by employing only (roughly) half the number of random choices.

## 1 Introduction

The problem of disseminating a piece of information in a given network is a fundamental problem in theoretical computer science. It has various applications, cf. Demers, Greene, Hauser, Irish, Larson, Shenker, Sturgis, Swinehart, and Terry [DGH[+]88] and Kempe, Dobra, and Gehrke [KDG03]. An important application is the maintenance of replicated databases.

The problem description is as follows. Assume that some new information is injected into the network (e.g., the updated address of some customer). We typically assume that in the beginning exactly one member of the network holds this piece of information (*rumor*) and we aim to spread the information to all other members of the network. The network is modeled by a graph, hence, its members are also referred to as nodes.

The dissemination of the information proceeds in synchronous rounds. Thus, in each round, each node which already has received the information in a previous iteration picks one of its neighbors and passes the information to it if the neighbor has not been informed already. This describes the PUSH version of the problem. The PULL version is symmetric. In each iteration, each node which has not yet received the information picks one of its neighbors and receives from it the information if the contacted neighbor owns it already. Of course, one can combine the two strategies to a PUSH-PULL protocol.

---
[*]Carola Winzen is a recipient of the Google Europe Fellowship in Randomized Algorithms, and this work is supported in part by this Google Fellowship.

It is an intensively studied question how to best transmit the information. Several algorithms for this problems have been developed. These protocols differ by how a node choses its neighbor to which it passes on the information or from which it requests the information, respectively. The three main quality criteria for the protocols are (i) *speed* or *broadcasting time*, i.e., number of rounds needed until all nodes are informed, (ii) *message complexity*, i.e., total number of times a node contacts one of its neighbors[1], and (iii) *robustness* against different types of node and/or transmission failures.

We note already here that in this work we are mainly interested in PUSH algorithms and that no PUSH protocol can broadcast the information faster than in $\lceil \log_2(n) \rceil$ rounds as the number of informed nodes at most doubles in each iteration. The minimal message complexity is $n-1$. It is easy to verify that there exist deterministic algorithms which meet these bounds. These algorithms, however, are not suited at all if one asks for robustness of the protocol. Therefore, several randomized algorithms for this problem have been developed.

The probably most intensively studied protocol are the random versions of the PUSH, PULL and PUSH-PULL protocols by Demers et al. [DGH+88], cf. also Feige, Peleg, Raghavan, and Upfal [FPRU90], Frieze and Grimmett [FG85], Karp, Schindelhauer, Shenker, and Vöcking [KSSV00]. In the random PUSH version, each informed node choses in each round the neighbor to contact independently and uniformly at random from the set of all its neighbors. We call this model the *fully random rumor spreading protocol*. Due to the random nature of the algorithm it is strongly robust against node and transmission failures, cf. [ES06], [HKP+04].

Inspired by the rotor router model for random walks on the graph (also referred to as Propp machine, cf. [PDDK96] and [HLM+08]), Doerr, Friedrich, and Sauerwald presented in [DFS08] a quasi-random analog of the fully random PUSH protocol. Despite the fact that their *(classical) quasi-random rumor spreading protocol* involves much less randomness, it performs at least as fast as the fully random protocol on many graph classes. At the same time it is reasonably robust, cf. [DFS09]. The protocol works as follows. Each node holds a cyclic list of its neighbors. Once a node becomes informed in some round $t$, it picks a starting point on the list uniformly at random and informs, in round $t+1$, the neighbor corresponding to that position. Thereafter, it continues informing its neighbors in the cyclic order of the list. We mention some results for the quasi-random protocol in the related work section.

---

[1]Sometimes, message complexity is defined as the total number of times the information is send from one node to one of its neighbors. This is in particular true for most papers dealing with PULL or PUSH-PULL protocols.

A major breakthrough in the studies of broadcasting algorithms is the recent work by Doerr and Fouz [DF10]. The authors present new quasi-random versions of the PUSH protocol which they call *hybrid models*. The algorithms are based on the classical quasi-random protocol by Doerr, Friedrich, and Sauerwald, but differ in two details which significantly speed up the random process of broadcasting. Firstly, they equip all nodes with the same list. This requires that all nodes have a unique ID – we feel that this is a reasonable assumption for most applications[2]. Secondly, they introduce the concept of *restarts*. That is, once a node tries to inform a node which has already been informed before, it will, in the next iteration, inform a random neighbor (chosen again uniformly at random from the list of all its neighbors) and then continues informing according to the cyclic list. Each node is allowed to perform up to $R-1$ such random restarts, $R$ being a parameter of the model. Doerr and Fouz show that this *quasi-random protocol with $R$ random calls* achieves for $R = \sqrt{\log(n)}$ an asymptotically optimal broadcasting time of $(1+o(1))\log_2(n)$. Furthermore, it has a message complexity of at most $n(R+1)$.

In this study we present an alternative set of quasi-random broadcasting algorithms which we call *direction-reversing quasi-random rumor spreading protocol*. The main idea is that the nodes store the information of their random decision and reverse their direction once they tried informing an already informed neighbor, i.e., they continue in acyclic direction of the list, starting from the neighbor "left" to their stored random decision. The precise model is given in Section 2. The direction-reversing model reduces the bounds obtained by Doerr and Fouz even further. Put differently, our protocols achieve the same runtime as the hybrid model by employing less random decisions. This shows that despite being asymptotically optimal, it is possible — even by simple means — to further decrease the runtime of the currently best known PUSH protocols.

## 1.1 Related work

The literature on the rumor spreading problem is huge. Therefore, we shall restrict ourselves to results on the complete graph. Frieze and Grimmett [FG85] proved that the fully random protocol informs all nodes within $(1+o_p(1))(\log_2(n) + \ln(n))$ iterations where $o_p(1)$ denotes a random variable which converges to 0 in probability[3]. Pittel [Pit87] improved this bound to $\log_2(n) + \ln(n) + O_p(1)$[4].

---

[2]Note also that Karp et al. [KSSV00] showed that any address-oblivious algorithm needs $\Omega(n \log \log n)$ rounds to inform all nodes in the network.

Angelopoulos, Doerr, Huber, and Panagiotou [ADHP09] showed that the upper bound by Frieze and Grimmett also applies to the classical quasi-random model. Later, Fountoulakis and Huber [FH09] showed that also Pittel's bound carries over. More precisely, they show that the classical quasi-random rumor spreading protocol informs, with high probability (w.h.p., i.e., with probability $1 - o(1)$), all vertices within $[\log_2(n) + \ln(n) - 4\ln\ln(n), \log_2(n) + \ln(n) + h(n)]$ iterations, $h = \omega(1)$ being a function of arbitrarily slow growth.

Finally, Doerr and Fouz achieve the above mentioned asymptotically optimal bounds for their hybrid model. More precisely, they show the following. If $\varepsilon > 0$ and $h = \omega(1)$, it holds w.h.p. that the hybrid model with $R$ random calls informs all nodes within $\log_2(n) + (1+\varepsilon)\ln(n)/R + R + h(n)$ iterations if $R \leq \sqrt{\ln(n)}$ and within $\log_2(n) + (2+\varepsilon)\sqrt{\ln(n)}$ for larger values of $R$. They also show that the result is tight in the sense that with probability $1 - \exp(-n^{\Theta(\varepsilon)})$ there exists at least one uninformed node after $\log_2(n) + (1-\varepsilon)\ln(n)/R + R/2$ iterations for $R \leq \sqrt{2(1-\varepsilon)\ln(n)}$ and after $\log_2(n) + \sqrt{(2-\varepsilon)\ln(n)}$ iterations for larger $R$.

## 1.2 Our Results

As mentioned above, we introduce a direction-reversing alternative to the protocol by Doerr and Fouz. We are able to show the following.

**Theorem 1.** *Let $h(n) = \omega(1)$ be a function of arbitrarily slow growth and let $\varepsilon > 0$ be an arbitrarily small constant. It holds with probability $1 - o(1)$ that the direction-reversing quasi-random broadcasting protocol with $R$ random calls per node informs all nodes within*

$$\log_2(n) + \lceil (1+\varepsilon)\ln(n)/(2R) \rceil + 2R + h(n)\,, \text{ if } R \leq \tfrac{1}{2}\sqrt{\ln(n)} \text{ and}$$
$$\log_2(n) + (2+\varepsilon)\sqrt{\ln(n)}\,, \text{ if } R > \tfrac{1}{2}\sqrt{\ln(n)}\,.$$

*iterations. The message complexity of this model is at most $n(2R+1)$.*

That is, for $R < \tfrac{1}{2}\sqrt{\ln(n)}$, our direction-reversing rumor spreading protocol achieves a constant factor improvement over the hybrid model. For $R = 1$ this improvement equals $\big((\ln(2)^{-1} + 1/2 + \varepsilon/2)\ln(n) + 2\big)/\big((\ln(2)^{-1} + 1 + \varepsilon)\ln(n) + 1\big) \approx .795$, i.e., when allowing the same dose of randomness, our model reduces the known broadcasting time of the hybrid model by 20%. Of course, at the same time this is a 20% improvement over the bound given in [FH09] for the classical quasi-random protocol. For $R = 2$ the advantage of our model over the hybrid model still is approximately 13% which translates to a 30% improvement over the classical quasi-random model.

---

[3] We say that a random variable $X_n$ converges to 0 in probability if for all $\varepsilon > 0$ it holds that $\Pr[|X_n| > \varepsilon] = o(1)$.

[4] Here we denote by $O_p(1)$ a random variable $X_n$ which satisfies with probability $1-o(1)$ that $|X_n| \leq h(n)$ where $h = \omega(1)$ is of arbirtrarily slow growth.

As discussed above, this result shows that although being asymptotically optimal, the broadcasting time achieved by the hybrid model can be improved even by simple ideas. However, we will see in Section 3 that the proof for Theorem 1 becomes more involved than its corresponding version for the hybrid model.

Similar to the hybrid model we are able to prove that the bounds given in Theorem 1 are essentially sharp.

**Theorem 2.** *For any $\epsilon > 0$ it holds with probability $1 - \exp(-n^{\Theta(n)})$ that the direction-reversing quasi-random broadcasting protocol with $R$ random calls per node is not finished within the following number of iterations*

$$\log_2(n) + (1-\varepsilon)\ln(n)/(2R) + R/2 \, , \text{ if } R \leq \sqrt{(1-\varepsilon)\ln(n)} \text{ and}$$
$$\log_2(n) + \sqrt{(1-\varepsilon)\ln(n)} \, , \text{ if } R \geq \sqrt{(1-\varepsilon)\ln(n)} \, .$$

The proof of Theorem 2 is deferred to the appendix. Essentially, it follows the lines of its equivalent version for the hybrid model in [DF10].

## 1.3 Disclaimer

The drawback of both the algorithms by Doerr and Fouz and our direction-reversing ones is the fact that it can only be applied to the complete graph. In alignment with Doerr and Fouz, however, we think that it is an interesting enough setting to experiment with new ideas. Note also that in this study we are only considering the speed and the message complexity of the protocols. Due to the random behavior of the models, however, we can expect that the protocols are reasonably robust as well.

## 2 The Direction-Reversing Model

We consider the complete undirected and labeled graph with $n$ vertices (which simultaneously also call nodes). Without loss of generality, we assume that the nodes are labeled by $0, \ldots, n-1$.

In the following, we shall describe an algorithm which disseminates some piece of information that has been newly injected into the network. Due to the symmetry of the problem we may assume that initially node 0 holds some piece of information ("the *rumor*") that shall be distributed to everyone in the network. As mentioned in the Introduction, we equip all nodes with the same list. Again easing readability we may assume that this is the trivial list $(0, \ldots, n-1)^5$. The information is transmitted along the edges, in both directions, and we assume that the broadcasting proceeds in rounds. However, we want to ensure that each node becomes informed by exactly one of its neighbors and therefore require that no two nodes try broadcasting their information at *exactly* the same time.

Throughout the paper, all node labels should be understood modulo $n$. That is, if we speak of node $i$ and $i$ happens to be larger or equal to $n$, we actually refer to node $i \mod n$ and we apply the same logic for negative values of $i$. By the same reasoning we have an cyclic understanding of the word "interval". That is, we call $I$ an interval if there exists an interval $I' = [i_1, \ldots, i_\ell] \subseteq \mathbb{Z}$ of integers such that $I = [i_1 \mod n, \ldots, i_\ell \mod n]$. By slight abuse of language we also call $I$ an interval even if it only consists of one single vertex $I = [i] = \{i\}$.

---

**Algorithm 1:** Routine of Vertex $v$ in the Direction-Reversing Broadcasting Protocol with $R$ Random Calls.

**1 for** $i = 1$ **to** $R$ **do** pick $v_i(v) \in [0, \ldots, n-1]$ uniformly at random;
**2** Copy $j \leftarrow v_i(v)$;
**3 while** $j$ *not informed* **do**
**4** $\quad$ pass information and update $j \leftarrow j + 1$;
**5** Update $j = v_i(v) - 1$;
**6 while** $j$ *not informed* **do**
**7** $\quad$ pass information and update $j \leftarrow j - 1$;

---

As mentioned above, our algorithm builds on the idea of Doerr and Fouz in [DF10]. Recall that their algorithm has two main ingredients which speed up the random process of broadcasting: identical lists for all nodes and random restarts. Our algorithm differs from the one in [DF10] in that we add the idea of reversing the direction of the broadcasting process. To this end, we propose the following algorithm.

Whenever a vertex $v$ becomes informed in round $t$, it follows the routine described in Algorithm 1. That is, it picks in round $t + 1$ a starting point $\nu_1(v) \in \{0, \ldots, n-1\}$ uniformly at random and informs it if it hasn't been informed already. From then on it informs in each subsequent iteration the next node in the cyclic order of the list until it tries informing a node which has been informed already. In this case, it reverses its direction and tries informing (in the next round) node $\nu_1(v) - 1$. From there it continues informing neighbors in the reversed direction of the cyclic list until it again tries to inform a node which has been informed already. If this happens, it makes a random restart and picks $\nu_2(v) \in \{0, \ldots, n-1\}$ uniformly at random and so on.

---

[5] Note that for readability purposes we assume that all nodes have *exactly* the same list. That is, each node is contained in its own list. It will be obvious for all protocols that self-containing lists are at most as fast as the protocols where nodes do not appear on their own list.

If $v$ makes the $i$-th random call in iteration $t_i(v)$ and the $(i+1)$-st random call in iteration $t_{i+1}(v)$ we call the iterations $[t_i(v), t_{i+1}(v) - 1]$ the *$i$-th trial* of node $v$.

A node stops informing other nodes after iteration $t_{\text{stop}}(v)$ if after the $R$-th random call and after having reversed its direction it again tries in iteration $t_{\text{stop}}(v)$ to inform an already informed node. Similar to the definition above we call the interval $[t_R(v), t_{\text{stop}}(v)]$ the *$R$-th trial* of vertex $v$.

By $\mathcal{I}_t$ we denote the set of all vertices which have received the rumor in any of the first $t$ rounds. Accordingly, we denote by $\mathcal{U}_t$ the set of vertices which, at the end of round $t$, have still not received the rumor, i.e., $\mathcal{U}_t = \{0, \ldots, n-1\} \backslash \mathcal{I}_t$. Node $v$ is said to be *newly informed* in round $t+1$ if it has received the rumor in the $t$-th iteration, i.e., $v \in \mathcal{I}_t \cap \mathcal{U}_{t-1}$. Thus, a node which is newly informed in round $t+1$ tries for the first time in round $t+1$ to inform one of its neighbors.

The *runtime* or *broadcasting time* of the algorithm is the number of iterations needed until all nodes in the network have received the information. Note that throughout the paper we simultaneously speak of iterations and rounds.

Before we analyze the protocol we need to make two more technical assumptions, both very similar to assumptions made in [DF10]. The first one concerns the beginning of the broadcasting process. We assume the following. The starting node, node 0, does not pick $\nu_1(0) \in \{0, \ldots, n-1\}$ uniformly at random but we fix $\nu_1(0) = 0$. For readability purposes we say that node 0 informed itself in round 0. Everything else in the first trial and all other trials proceed as described by Routine 1. In particular, we have that in round 1 of the algorithm node 0 informs node 1. The second assumption deals with the timing of the broadcasting within one iteration. Recall that we assume that no two nodes call their neighbors at *exactly* the same time. To avoid the event that a random call delays the time in which a particular node becomes informed (cf. Observation 3 in the appendix), we assume that within one iteration the random calls are performed latest. I.e., first those nodes call their neighbors which have informed an uninformed node in the previous round or are reversing their direction. Thereafter all newly informed and restarting nodes try informing their neighbors[6].

---

[6]Although in this document we do not elaborate this idea further, we shall mention that the second assumption can be avoided by some more technical effort in the analysis of Theorem 1.

# 3 Proof of Theorem 1

Although the proof of Theorem 1 will be similar to the corresponding one for the hybrid model, it requires some more technical effort, most of which seems to be obviously true but will require, however, careful proofs.

We begin with some observations. The first one is due to the fact that in the classical quasi-random rumor spreading the concept of restarts does not exist and thus, when equipped with identical lists, the nodes continue calling already informed neighbors once they have tried to inform a node that has been informed already.

**Observation 1.** *The direction-reversing quasi-random broadcasting protocol with $R$ calls is always at least as fast as the classical quasi-random model with identical lists.*

The next observation is due to the fact that all nodes have the same list $(0, \ldots, n-1)$ of their neighbors and that all nodes follow this list in cyclic order (before eventually reversing the direction).

**Observation 2.** *Let $u, i \leq n$ such that node $u - i$ is has received the rumor in the first $t$ rounds, i.e., $u - i \in \mathcal{I}_t$. It takes at most $i$ additional rounds until $u$ becomes informed. In particular, $u \in \mathcal{I}_{t+i}$.*

If we consider the corresponding reversed situation (cf. situation in Lemma 3), Observation 2 is not that obvious any more. Even worse, it becomes false if one does not require that, within one round of the algorithm, the running processes inform their neighbors first and only thereafter newly informed and restarting nodes are allowed to call their neighbors. We give an example for this phenomenon in the appendix, cf. Observation 3.

**Lemma 3.** *Let $u, i \leq n$ such that $u + i \in \mathcal{I}_t$. Assume that the node which informed $u + i$ has reversed its direction already, i.e., it tries to inform $u + i - 1$ in the next iteration, iteration $t + 1$. Then it holds that $u \in \mathcal{I}_{t+i}$.*

*Proof.* We prove the claim via an induction over the distance $i$ from $u$ to $u + i$. Clearly, if $i = 1$ there is nothing to show as the node which informed $u + 1$ has reversed its direction already and will try to inform node $u$ in iteration $t + 1$. Hence, $u$ becomes informed in the $(t+1)$-st iteration, if it wasn't already — either through the node which informed $u + 1$ or another node.

Therefore, let us assume that $i \geq 2$ and that the claim is true for all smaller values of $i$. We first consider the case where $[u, \ldots, u+i-1] \subseteq \mathcal{U}_t$, i.e., all nodes between $u$ and $u + i - 1$ are uninformed. Let us denote by $w$ the node which informed vertex $u + i$. Note that by definition, it has reversed its direction already. Thus, if in iterations $t+1, \ldots, t+i$ no other node tries to inform any of the vertices $u+i-1, u+i-2, \ldots, u+1, u$, node $w$ will call node $u$ in iteration $t + i$ and thus, $u \in \mathcal{I}_{t+i}$.

Therefore, let us consider the case where there exists a node $w'$ which successfully tries to inform some node $u + i' > u$ in iteration $t + t'$. We consider the smallest such $t'$ and smallest such $i'$. Since $[u, \ldots, u+i-1] \subseteq \mathcal{U}_t$ holds, node $w'$ must have performed a random call in iteration $t+t'$. In this same iteration node $w$ called vertex $u + i - t'$. Since the random calls are performed only after the running processes have informed their neighbors and since node $w'$ successfully informed $u + i'$, it must hold that $i' < i - t'$ ((**i**)).

Set $U_1 := [u, \ldots, u + i' - 1]$, $U_2 := [u + i' + 1, \ldots, u + i - t' - 1]$, and let $u_2 := |U_2| = i - t' - i' - 1$, the length of $U_2$. By Observation 2 it holds that all nodes in $[u + i' + 1, u + i' + \lceil \frac{u_2}{2} \rceil]$ are informed within the next $\lceil \frac{u_2}{2} \rceil$ iterations. Furthermore we can apply the induction hypothesis to the nodes in $[u+i'+\lceil \frac{u_2}{2}+1 \rceil, \ldots, u+i-t-1]$ and node $u+i-t'$. Recall that the latter one has been informed in iteration $t + t'$ by node $w$. Therefore, all nodes in $[u + i' + \lceil \frac{u_2}{2} + 1 \rceil, \ldots, u + i - t - 1]$, too, are informed after an additional $\lceil \frac{u_2}{2} \rceil$ number of iterations.

Summarizing these statements we have that, at the end of iteration $t + t' + \lceil \frac{u_2}{2} \rceil$ all nodes in $U_2$ are informed. Hence, the latest possible round in which node $w'$ reverses its direction is iteration $t+t'+\lceil \frac{u_2}{2} \rceil +1$, meaning that $w'$ calls node $u+i'-1$ in iteration $t+t'+\lceil \frac{u_2}{2} \rceil +2$, at latest. By again applying the induction hypothesis, this time to node $u$ and node $u + i'$ in iteration $t+t'+\lceil \frac{u_2}{2} \rceil +1$, we get that $u$ is informed after iteration $t+t'+\lceil \frac{u_2}{2} \rceil +1+i' = t+t'+\lceil \frac{i-t'-i'-1}{2} \rceil +1+i'$. Applying inequality ((**i**)) it is easy to verify that this expression can be bounded from above by $t + i$.

Lastly, let us remark that the case where $[u, \ldots, u + i - 1] \not\subseteq \mathcal{U}_t$ follows by exactly the same reasoning as above. If node $u + i' \in \mathcal{I}_t$ there exists a vertex $w'$ which informed $i'$ in some iteration $\leq t$. If we chose $i'$ to be maximal we conclude as above that node $w'$ reverses its direction in iteration $\lceil (i-i'-1)/2 \rceil + 1$, at latest. Again by applying the induction hypothesis we get that node $u$ will be informed at the end of round $t+\lceil (i-i'-1)/2 \rceil +1+i'$. Since $i' < i$, it is again easy to verify that this expression can be bounded by $t + i$. □

We immediately gain the following two statements.

**Corollary 4.** *Let $U = [u_1, \ldots, u_\ell] \subseteq \mathcal{U}_t$ be an interval of uninformed vertices. Assume that $U$ is maximal, i.e., $u_1 - 1$ and $u_\ell + 1$ are informed already. Furthermore, assume that the node which informed $u_\ell + 1$ will inform $u_\ell$ in round $t + 1$, i.e., it has tried in a previous round of the same trial to inform an already informed node and, consequently, has reversed its direction.*

*Then it takes at most $\lceil |U|/2 \rceil$ additional rounds until all nodes of $U$ have received the rumor.*

**Corollary 5.** *If after $t$ iterations of the direction-reversing quasi-random rumor spreading protocol $U = [u_1, \ldots, u_\ell]$ is the only remaining interval of uninformed vertices (i.e., $\mathcal{U}_t = U$) and the nodes which informed $u_1 - 1$ and $u_\ell + 1$, respectively, are not identical, it takes at most $\lceil(|U|-1)/2\rceil + 1$ many rounds until all nodes are informed.*

*Proof.* Either the node which informed $u_\ell + 1$ has reversed its direction already in which case it follows from Corollary 4 that all nodes (in $U$ and thus, all nodes) are informed within $\lceil |U|/2 \rceil$ iterations. Otherwise, the node reverses its direction after the $(t+1)$-st round and again we can apply Corollary 4, this time to $[u_2, \ldots, u_\ell]$ since the node which informed $u_1 - 1$ calls and informs node $u_1$ in the $(t+1)$-st iteration. □

Corollary 5 is particularly useful in combination with the following lemma.

**Lemma 6.** *If after $t \geq 2$ iterations of the direction-reversing quasi-random rumor spreading protocol $U = [u_1, \ldots, u_\ell]$ is the only remaining interval of uninformed vertices, then, with high probability, the node which informed $u_1 - 1$ and the one which informed $u_\ell + 1$ are different.*

*Proof.* Let us denote by $w_-$ the node which informed $u_1 - 1$ and by $w_+$ the one which informed $u_\ell + 1$. We first observe that $w_- = w_+$ can hold only, if the only node which has successfully informed other nodes is the starting node, node 0. That is, it must hold that $w_- = w_+ = 0$. But this implies that node 1, which has been informed in the first round, has chosen $\nu_1(1) \in \{1, 2\}$, i.e., it tried to informed itself or node 2 which, in the beginning of the second iteration, has been informed by node 0. As this happens with probability $2n^{-1}$, the claim follows. □

Next, we need another lemma which will play a key role in the proof of Theorem 1.

**Lemma 7.** *Let, at the end of the $t$-th iteration, $(U_1, U_2)$ be two consecutive intervals of uninformed vertices (in the cyclic order of the list $(0, \ldots, n-1)$). If $|U_1| + |U_2|$ is maximal, then $\lceil(|U_1|+|U_2|+1)/2\rceil$ is an upper bound for the remaining number of iterations until all nodes in the network have received the rumor.*

*Proof.* Let $u \in \mathcal{U}_t$. Let $U_1 = [u_1^{(1)}, \ldots, u_{\ell_1}^{(1)}] \subseteq \mathcal{U}_t$ be the *maximal interval* of uninformed vertices with $u \in U_1$. That is, all nodes in $U_1$ are uninformed, whereas the two adjacent nodes $u_1^{(1)} - 1$ and $u_{\ell_1}^{(1)} + 1$ have already been informed.

Let $U_2 = [u_1^{(2)}, \ldots, u_{\ell_2}^{(2)}]$ be the maximal interval of uninformed vertices such that all nodes between $U_1$ and $U_2$ are informed.

If $\ell_1 \leq \ell_2$, it takes, according to Observation 2, at most $\ell_1 \leq (\ell_1 + \ell_2)/2$ additional iterations until $u$ becomes informed. Hence, we can assume without loss of generality that $\ell_1 > \ell_2$.

Again by Observation 2 it takes at most $\ell_2$ rounds until all nodes in $U_2$ are informed. Then, in the worst case, it may take one additional round until the node which informed $u^{(1)}_{\ell_1}+1$ (in some iteration $\leq t$) reverses its direction and tries to inform node $u^{(1)}_{\ell_1}$. That is, at the end of the $(\ell_2+1)$-st iteration we are in the situation of Lemma 3, yielding that $u \in \mathcal{I}_{t+\ell_2+1+(u^{(1)}_{\ell_1}-u)}$. Furthermore, it follows from Observation 2 that all nodes $u^{(1)}_1, \ldots, u^{(1)}_{\ell_2+1} \in \mathcal{I}_{t+\ell_2+1}$ and, by the same Observation we conclude that $u \in \mathcal{I}_{t+\ell_2+1+(u-u^{(1)}_{\ell_2+1})}$.

Putting everything together we have that $u$ will have received the rumor at the end of iteration $\min\{t+\ell_2+1+(u^{(1)}_{\ell_1}-u), t+\ell_2+1+(u-u^{(1)}_{\ell_2+1})\}$. Now, $\min\{u^{(1)}_{\ell_1}-u, u-u^{(1)}_{\ell_2+1}\} \leq \lceil(\ell_1-(\ell_2+1))/2\rceil$ and thus, starting from round $t$, it takes at most $\ell_2+1+\lceil(\ell_1-(\ell_2+1))/2\rceil = \lceil(\ell_1+\ell_2+1)/2\rceil$ additional iterations until $u$ is informed. $\square$

Finally, we need one last observation. In the proof of Theorem 1 we will make use of the so-called *delaying technique*. This technique has proven useful in the analysis of (quasi-)random rumor spreading protocols (cf. [DFS08],[ADHP09]). We say that a vertex $v$ is *delayed*, if it stops, for a given number of rounds, informing other nodes after it has tried for the second time to inform a node which had already been informed before. That is, instead of making the second random call in $t_2(v)$ the vertex stops informing its neighbors after the first trial. We say that the direction-reversing protocol is delayed if we delay all (informed) nodes.

By Observation 2 and Lemma 3 the following is true.

**Corollary 8.** *The delayed protocol cannot be faster than the original direction-reversing protocol.*

The proof of Theorem 1 can now be conducted similarly as in [DF10].

*Proof of Theorem 1.* We split the process into three phases. The first one lasts for $\log_2(n)+h(n)$ rounds. We assume that in this first phase the nodes are delayed (i.e, after having tried for the first time to inform a node which has been informed before, they stop informing other nodes for the remainder of the first phase). By Corollary 8 it holds that our (non-delayed) direction-reversing quasi-random protocol is at least as fast as the delayed protocol which in turn is, by Observation 1, at least as fast as the classical quasi-random protocol with identical lists. Lastly, it holds by definition of the classical quasi-random protocol (which assumes worst-case lists), that the classical quasi-random protocol with identical lists is at least as fast as the classical quasi-random protocol. Summarizing these observations we have that

classical quasi-random protocol
$$\leq \text{classical quasi-random protocol with identical lists}$$
$$\leq \text{delayed direction-reversing quasi-random protocol}$$
$$\leq \text{direction-reversing quasi-random protocol},$$

where $A \leq B$ indicates that protocol $B$ is at least as fast as protocol $A$. It has been shown by Fountoulakis and Huber [FH09] for the classical quasi-random model that with probability $1 - o(1)$ it holds at the end of the first phase that $(1 - \epsilon)n$ nodes are informed, for an arbitrarily small constant $\epsilon$.

The second phase lasts for $2R$ rounds. Note that in the second phase each of the $(1 - \epsilon)n$ informed vertices makes a random call after exactly 2 iterations unless it has done $R$ random calls already or it informs a node which has not been informed before. In the latter case, the newly informed node makes a random call in the next iterations. That is, the number of random calls at the end of the second phase, including the random calls from the first phase, is at least $(1 - \epsilon)nR$.

We need to show that after the second phase it takes at most $\lceil (1 + 2\epsilon) \ln(n)/(2R) \rceil$ iterations until all nodes in the network have been informed. The claim then follows from setting $\epsilon = \varepsilon/2$.

We show the following.

**(A)** Each interval $U = [u_1, \ldots, u_\ell]$ of uninformed vertices has length less than $(1 + 2\epsilon) \ln(n)/R$, w.h.p.

**(B)** If there are two consecutive intervals $(U_1, U_2) = ([u_1^{(1)}, \ldots, u_{\ell_1}^{(1)}], [u_1^{(2)}, \ldots, u_{\ell_2}^{(2)}])$ of uninformed vertices such that the node which informed $u_{\ell_1}^{(1)} + 1$ is the same as the one which informed $u_1^{(2)} - 1$, it holds w.h.p. that $\ell_1 + \ell_2 < (1 + 2\epsilon) \ln(n)/R$.

From **(A)** and **(B)** we derive the claim as follows.

If at the end of the second phase there is only one interval $U$ of uninformed vertices left, it follows from **(A)** that it has length at most $(1 + 2\epsilon) \ln(n)/R - 1$ and by Corollary 5 and Lemma 6 we obtain that, with high probability, all nodes in $U$ are informed within the next $\lceil (|U| - 1)/2 \rceil + 1 \leq \lceil ((1 + 2\epsilon) \ln(n)/R - 2)/2 \rceil + 1 = \lceil (1 + 2\epsilon) \ln(n)/(2R) \rceil$ iterations.

If at the end of the second phase there exist more than only one interval of uninformed vertices, it holds for each maximal one of them, $U_1 = [u_1^{(1)}, \ldots, u_{\ell_1}^{(1)}]$, that either the node which informed node $u_{\ell_1}^{(1)} + 1$ has (i) reversed its direction already, (ii) it will do so in the next iteration, or, (iii) it is the same one which informed $u_1^{(2)} - 1$ if $U_2 = [u_1^{(2)}, u_{\ell_2}^{(2)}]$ denotes the next interval of uninformed vertices in the cyclic order of the list $(0, \ldots, n - 1)$.

In the first case, (i), we can apply Corollary 4 and obtain that all nodes in $U_1$ are informed within $\lceil |U_1|/2 \rceil$ subsequent iterations. This term can, by **(A)**, be bounded from above by $\lceil (1+2\epsilon)\ln(n)/(2R) \rceil$.

In the second case, (ii), all nodes in $U_1$ are informed after an additional $\lceil (|U_1| - 1)/2 \rceil + 1$ iterations, again by Corollary 4. We again apply **(A)** to bound this expression by $\lceil ((1+2\epsilon)\ln(n)/R - 2)/2 \rceil + 1 = \lceil (1+2\epsilon)\ln(n)/(2R) \rceil$.

In the third case, (iii), we apply Lemma 7 to obtain that all nodes in $U_1$ and $U_2$ are informed within $\lceil (|U_1| + |U_2| + 1)/2 \rceil$ iterations. By expression **(B)** we can bound this term from above by $\lceil ((1+2\epsilon)\ln(n)/R - 1 + 1)/2 \rceil = \lceil (1+2\epsilon)\ln(n)/(2R) \rceil$.

This shows that from **(A)** and **(B)** we can derive that all nodes which have not been informed in the first two phases will become informed within the $\lceil (1+2\epsilon)\ln(n)/(2R) \rceil$ subsequent iterations, with high probability. By taking a simple union bound over the error terms of the three different phases we conclude that the total broadcasting time is at most $\log_2(n) + h(n) + 2R + \lceil (1+2\epsilon)\ln(n)/(2R) \rceil$, with high probability.

To derive **(A)** we first compute the probability that $(1+2\epsilon)\ln(n)/R$ particular nodes are not being hit by any one of the random calls. This clearly is an upper bound for the probability that the nodes are still uninformed at the end of the second phase. Since there are at least $(1-\epsilon)nR$ random calls in the first two phases, we bound this probability by

$$\left(1 - \frac{(1+2\epsilon)\ln(n)}{nR}\right)^{(1-\epsilon)nR} \leq \exp\left(-(1-\epsilon)(1+2\epsilon)\ln(n)\right) = n^{-(1+\epsilon-2\epsilon^2)} = n^{-(1+\epsilon')}$$

for $\epsilon' := \epsilon - 2\epsilon^2$ (which clearly is $> 0$ for sufficiently small $\epsilon$).

Next we apply a simple union bound to obtain an upper bound for the probability that there exists an interval of uninformed vertices $U = [u_1, \ldots, u_\ell]$ of length at least $(1+2\epsilon)\ln(n)/R$. Note that there are less than $n$ possible positions for $u_1$ and thus, the probability that such an interval $U$ exists is less than $n \cdot n^{-(1+\epsilon')} = o(1)$. This shows **(A)**.

It remains to prove **(B)**. To this end let $(U_1, U_2) = ([u_1^{(1)}, \ldots, u_{\ell_1}^{(1)}], [u_1^{(2)}, \ldots, u_{\ell_2}^{(2)}])$ be two consecutive intervals of uninformed vertices such that the node which informed $u_{\ell_1}^{(1)} + 1$ is the same one as the one which informed $u_1^{(2)} - 1$.

We note the following.

1. There are less than $n$ possible positions for $u_1^{(1)}$.

2. According to **(A)**, $U_1$ has length less than $(1+2\epsilon)\ln(n)/R$. That is, there are less than $(1+2\epsilon)\ln(n)/R$ possibilities for the length $\ell_1$ of $U_1$.

3. Between $U_1$ and $U_2$, there are at most $\log_2(n) + h(n) + 2R$ informed nodes. This follows from the fact that the node which informed $u_{\ell_1}^{(1)} + 1$ also informed $u_1^{(2)} - 1$ and that at the end of the second phase the total number of iteration elapsed is $\log_2(n) + h(n) + 2R$.

Thus, we can bound the probability that there exist a pair $(U_1, U_2)$ as in **(B)** from above by $n \cdot \ell_1 \cdot (\log_2(n) + h(n) + 2R) \cdot (1 - \frac{\ell_1 + \ell_2}{n})^{(1-\epsilon)nR}$. For $\ell_1 + \ell_2 \geq (1 + 2\epsilon) \ln(n)/R$ we can bound this expression from above by

$$n \frac{(1+2\epsilon)\ln(n)}{R}(\log_2(n) + h(n) + 2R)\left(1 - \frac{(1+2\epsilon)\ln(n)/R}{n}\right)^{(1-\epsilon)nR}$$
$$\leq n \frac{(1+2\epsilon)\ln(n)}{R}(\log_2(n) + h(n) + 2R)n^{-(1+\epsilon')}$$
$$= o(1).$$

That is, w.h.p., such a pair $(U_1, U_2)$ does not exist. This shows **(B)**.

Lastly, let us remark that the message complexity of the direction-reversing model is at most $n(2R+1)$ as there are at most $n-1$ successful calls (those leading to a newly informed vertex) and at most $n \cdot 2R$ unsuccessful ones until all nodes stop informing their neighbors. □

# Appendix

## Proof of Theorem 2

Whereas the proof of Theorem 1, the upper bound, required some more technical effort as the proof for the original quasi-random protocol with restarts given by Doerr and Fouz in [DF10], their proof for the lower bound almost carries over. We shall, however, provide a detailed proof for the sake of completeness. Note that accounting for the reversal of directions does mainly account for different constants and for a slightly different definition of *unaffectedness* (see below).

*Proof of Theorem 2.* If $R \leq \sqrt{(1-\varepsilon)\ln(n)}$ let $\Delta := (1-\varepsilon)\ln(n)/(2R) + R/2$ and let $\Delta := \sqrt{(1-\varepsilon)\ln(n)}$ otherwise. Furthermore, let $T := \log_2(n) + \Delta$. We need to prove that at the end of the $T$-th phase there exists at least one uninformed node, with probability at least $1 - \exp\left(-n^{\Theta(\varepsilon)}\right)$. The outline of the proof is as follows. First, we consider one specific vertex $u \geq 2T$ and show that the probability that $u$ remains uninformed within the first $T$ iterations of the algorithm can be bounded from below by $n^{-1+\Theta(\varepsilon)}$. We then conclude the proof by the following observations.

We first chose $k := \lfloor n/(2T+1) \rfloor - 1$ nodes $u_1, \ldots, u_k$ which have distance at least $2T+1$ from each other and from the starting node 0. We then denote by $E_i$ the event that node $u_i$ is informed after the $T$-th iteration. The events $E_1, \ldots, E_k$ are negatively correlated. This yields

$$\Pr[\text{ all } n \text{ nodes are informed within the first } T \text{ rounds}] \leq \Pr[\bigwedge_{i=1}^{k} E_i]$$

$$\leq \prod_{i=1}^{k} \Pr[E_i] \leq \left(1 - n^{-1+\Theta(\varepsilon)}\right)^k \leq \exp\left(-n^{\Theta(\varepsilon)}\right).$$

Thus, it suffices to show that the probability that one specific vertex $u \geq 2T$ does not get informed within the first $T$ rounds can be bounded from below by $n^{-1+\Theta(\varepsilon)}$. To this end we set $\epsilon := \varepsilon/8$ and divide the first $T$ rounds into the following three phases. The first phase consists of rounds 1 through $(1-\epsilon)\log_2(n)$, the second of of rounds $(1-\epsilon)\log_2(n) + 1$ through $\log_2(n)$ and the third one of the remaining rounds $\log_2(n) + 1$ through $T$.

We say that node $u$ remains *unaffected* by a random call in iteration $i \leq T$ if the random call avoids $[u - (T-i), u + T - i - 1]$, i.e., it does not call node $u$ nor the $T-i$ nodes "to the left" of $u$, nor the $T-i-1$ nodes "to the right" of $u$. If a node $u$ is unaffected by a random call, it will not become informed through the process initiated by that random call. Thus, we are interested in finding lower bounds for the probability that $u$ remains unaffected throughout all three phases. We say that $u$ is *affected* by a random call if it is not unaffected. Note that the fact that a random call in round $i$ calls a node in $[u+1, u+T-i-1]$ does not necessarily imply that $u$ becomes informed within the first $T$ rounds of the algorithm. We thus adopt a worst-case view in this proof.

Let us begin with the first phase. As the number of informed nodes at most doubles in each iteration, there are at most $n^{1-\epsilon}$ nodes which become informed in this first phase. Note that the number of random calls in the first phase can be bounded by the maximal number of newly informed vertices. This is due to the fact that each vertex which successfully informed a vertex in the previous iteration does not make a random call in the current one. Thus, there are at most $n^{1-\epsilon}$ random calls in the first phase and the probability that $u$ is affected by a particular one of these can be bounded from above by $2T/n$. By applying a union bound we can bound the probability that $u$ is affected in the first phase from above by

$$n^{1-\epsilon} 2T/n \leq 2n^{-\epsilon}(\log_2(n) + (1-\varepsilon)\ln(n)/2R + R/2) < 6n^{-\epsilon}\log_2(n),$$

with some room to spare. This yields

$$\Pr[u \text{ remains unaffected in the first phase}] \geq 1 - 6n^{-\epsilon}\log_2(n) = 1 - o(1) \tag{1}$$

By the same reasoning as above there are at most $n$ random calls in the second phase. The probability of a particular one of these to affect $u$ can be bounded from above by $2(\epsilon\log_2(n) + \Delta)/n$. As the random calls are independent of each other, we can bound the probability that $u$ remains unaffected in the second phase from below by

$$\Pr[u \text{ remains unaffected in the second phase}]$$
$$\geq \left(1 - 2(\epsilon\log_2(n) + \Delta)/n\right)^n$$
$$= (1 - o(1))\exp(-2(\epsilon\log_2(n) + \Delta))$$
$$\geq (1 - o(1))\exp(-4\epsilon\log_2(n)). \tag{2}$$

To bound the probability that $u$ remains unaffected in the third phase, we first observe that each of the $n$ nodes has at most $m := \min\{R, \Delta\}$ random calls to make. The probability that a random call in the $(\log_2(n) + i)$-th round affects $u$ can be bounded from above by $2(\Delta - i)/n$. We apply the inequality $1 - x > \exp(-x - x^2)$ valid for all $x \leq 1/2$ to obtain

$$\Pr[u \text{ remains unaffected in the third phase}] \geq \prod_{i=1}^{m}\left(1 - 2\tfrac{\Delta-i}{n}\right)^n$$
$$\geq \exp\left(-\sum_{i=1}^{m} 2(\Delta - i) + 4\tfrac{(\Delta-i)^2}{n}\right) = (1 - o(1))\exp\left(-2\sum_{i=1}^{m}(\Delta - i)\right)$$
$$= (1 - o(1))\exp\left(-2m\Delta + m(m+1)\right).$$

If $m = R$, then $R \leq \sqrt{(1-\varepsilon)\ln(n)}$ and thus,

Pr[$u$ remains unaffected in the third phase]
$$\geq (1 - o(1))\exp\big(-2R((1-\varepsilon)\ln(n)/(2R) + R/2) + R^2 + R\big)$$
$$= (1 - o(1))\exp\big(-(1-\varepsilon)\ln(n) + R\big)$$
$$= (1 - o(1))n^{-(1-\varepsilon)}.$$

Similarly, in case $m = \Delta$ we have that $R \geq \sqrt{(1-\varepsilon)\ln(n)}$ and $\Delta = \sqrt{(1-\varepsilon)\ln(n)}$. We thus obtain

Pr[$u$ remains unaffected in the third phase]
$$\geq (1 - o(1))\exp\big(-2\Delta^2 + \Delta^2 + \Delta\big)$$
$$= (1 - o(1))n^{-(1-\varepsilon)}.$$

Hence we can conclude for both cases that

Pr[$u$ remains unaffected in the third phase] $\geq (1 - o(1))n^{-(1-\varepsilon)}.$ (3)

Combining equations (1),(2), and(3) we obtain for $\epsilon = \varepsilon/8$ that

Pr[$u$ remains uninformed in the first $T$ rounds ]
$$\geq (1 - o(1)) \cdot \exp(-4\epsilon \log_2(n)) \cdot n^{-(1-\varepsilon)}$$
$$= (1 - o(1))n^{-1+\Theta(\varepsilon)}.$$

□

### Note Concerning the Timing of Random Calls

In the definition of the direction-reversing quasi-random rumor spreading protocol with restarts we required that within one round the first nodes to contact their neighbors are the ones which do not perform a random call in that same iteration. The following observation shows why this assumption is needed.

**Observation 3.** *If we omit the requirement given above, the delayed direction-reversing quasi-random rumor spreading protocol can be faster than the non-delayed version. Also, Lemma 3 must not hold in this situation.*

*Proof.* We consider the nodes in some interval $[i, i+1] \subseteq \mathcal{U}_t$ and we assume the following situation. In round $t + 1$ node $v$, which has already reversed its direction, successfully informs node $i + 1$. It then tries to inform node $i$ in round $t + 2$ and thus, $i\mathcal{I}_{t+2}$.

Assume now that there exist another node $r$ which makes a random call and tries informing node $i + 1$ in round $t + 1$, too. In our model, node $i + 1$ will have been informed already by node $v$ and thus, again, $i \in \mathcal{I}_{t+2}$ (it will actually be called by both $v$ and $r$ in the $(t+2)$-nd round).

If we do not assume that the random calls happen after all *running processes* have tried informing their neighbors, it may happen that node $i+1$ gets informed through node $r$. In this case, node $v$ will receive from node $i+1$ the information that it has been informed already. and it will either make a random restart (if it has performed less than $R$ random calls before the $t$-th iteration) or it totally stops informing other nodes. Node $r$ will try to inform node $i+2$ in the $(t+2)$-nd round and only thereafter reverses its direction. Hence it may take until the $(t+3)$-rd iteration until the $i$-th node becomes informed. □